\documentclass[a4paper,fleqn]{cas-sc}

\usepackage[authoryear,longnamesfirst]{natbib}
\usepackage{caption}
\usepackage{colortbl}
\usepackage{arydshln}

\usepackage{makecell}
\usepackage{url}
\usepackage{amsfonts}
\usepackage{amsmath}
\usepackage{multirow}
\usepackage{booktabs}
\usepackage{pifont}

\def\tsc#1{\csdef{#1}{\textsc{\lowercase{#1}}\xspace}}
\tsc{WGM}
\tsc{QE}

\begin{document}
\let\WriteBookmarks\relax
\def\floatpagepagefraction{1}
\def\textpagefraction{.001}

\shorttitle{A Dual-Branch Parallel Network for Speech Enhancement and Restoration}    

\shortauthors{D. Yang et al.}  

\title [mode = title]{A Dual-Branch Parallel Network for Speech Enhancement and Restoration}  



%





\author[1]{Da-Hee Yang}[type=editor, orcid=0009-0008-6253-2543,]
\author[2]{Dail Kim}
\author[1,2]{Joon-Hyuk Chang\footnote{Corresponding author. Email: jchang@hanyang.ac.kr}}
\author[3]{Jeonghwan Choi}
\author[3]{Han-Gil Moon}

\affiliation[1]{organization={Department of Electronic Engineering, Hanyang University},
            city={Seoul},
            postcode={04763}, 
            country={Republic of Korea}}
\affiliation[2]{organization={Department of Artificial Intelligence, Hanyang University},
            city={Seoul},
            postcode={04763}, 
            country={Republic of Korea}}
\affiliation[3]{organization={Samsung Electronics},
            city={Suwon},
            postcode={04763}, 
            country={Republic of Korea}}







\begin{abstract}
We present a novel general speech restoration model, \textbf{DBP-Net} (dual-branch parallel network), designed to effectively handle complex real-world distortions including noise, reverberation, and bandwidth degradation. Unlike prior approaches that rely on a single processing path or separate models for enhancement and restoration, DBP-Net introduces a unified architecture with \textit{dual parallel branches}—a masking-based branch for distortion suppression and a mapping-based branch for spectrum reconstruction. 
A key innovation behind DBP-Net lies in the \textit{parameter sharing} between the two branches and a \textit{cross-branch skip fusion}, where the output of the masking branch is explicitly fused into the mapping branch. This design enables DBP-Net to simultaneously leverage complementary learning strategies—suppression and generation—within a lightweight framework. Experimental results show that DBP-Net significantly outperforms existing baselines in comprehensive speech restoration tasks while maintaining a compact model size. These findings suggest that DBP-Net offers an effective and scalable solution for unified speech enhancement and restoration in diverse distortion scenarios.
\end{abstract}

\begin{highlights}
\item We propose DBP-Net, a dual-branch parallel network for general speech restoration under complex distortions such as noise, reverberation, and bandwidth reduction.
\item DBP-Net integrates suppression and generation pathways, with a learnable skip-fusion mechanism to effectively balance noise removal and bandwidth restoration.
\item The model employs a lightweight two-stage conformer backbone that captures both temporal and spectral dependencies for efficient speech representation.
\item Experimental results demonstrate that DBP-Net significantly outperforms existing baselines across perceptual quality and spectral distortion metrics.
\item DBP-Net achieves superior performance with a substantially smaller model size, providing an efficient yet powerful solution for general speech restoration.
\end{highlights}


\maketitle
\section{Introduction}
In real-world acoustic environments, speech signals are frequently degraded by a combination of distortions such as background noise, reverberation, and bandwidth limitations. While extensive research has been conducted on speech enhancement and bandwidth extension (BWE), most prior work has focused on single distortion types—such as denoising \cite{dnnse, cnnse, crnse, convtas, demucs, yangse, saleem2025ctse, zhang2025lretunet, chen2025ddp}, dereverberation \cite{dnnreverb, dereverb0, dereverb1}, or bandwidth extension \cite{dnnbwe, tunet}. However, in practical scenarios, these distortions often co-occur, rendering single-purpose systems insufficient.

Recent progress in general speech restoration has resulted in models that address multiple types of distortions, such as noise, reverberation, and bandwidth reduction \cite{liu2022voicefixer, byun2023empirical, kim2023hd, universal, universepp}. While these approaches mark important progress, they often rely on either task-specific modules or stage-wise processing pipelines, or require significantly larger model capacities, especially in the case of generative models. A key challenge in this domain lies in bridging the fundamentally different requirements of suppression (e.g., noise and reverberation removal) and reconstruction (e.g., bandwidth extension). Masking-based approaches are effective for suppression, while mapping-based generation is essential for reconstruction. Most existing discriminative models specialize in one of the two, limiting their ability to efficiently handle both within a unified framework.

To address these limitations, we propose DBP-Net, a dual-branch parallel network that explicitly models the two distinct tasks—enhancement and restoration—within a unified architecture. The model features two parallel branches: one masking-based branch focused on suppression, and one mapping-based branch designed for spectral reconstruction. Crucially, both branches share parameters and are connected via a cross-branch skip fusion, enabling the mapping-based branch to utilize intermediate representations from the masking branch. This structure encourages complementary learning and allows the network to adaptively balance suppression and reconstruction, even in complex distortion scenarios.
The novelty of DBP-Net lies in its architectural unification of two distinct learning paradigms through explicit branch collaboration and parameter sharing. Unlike prior approaches that either treat distortions separately or rely on sequential stages, DBP-Net enables simultaneous, efficient, and interpretable handling of multiple distortion types.

Experimental results demonstrate that DBP-Net achieves strong performance across various general speech restoration benchmarks while maintaining a low parameter count. Detailed analyses of its architecture and performance are provided in Section~\ref{sec:proposed} and Section~\ref{sec:experiments}, respectively.

\begin{figure*}[]
  \centering
  \includegraphics[width=\linewidth]{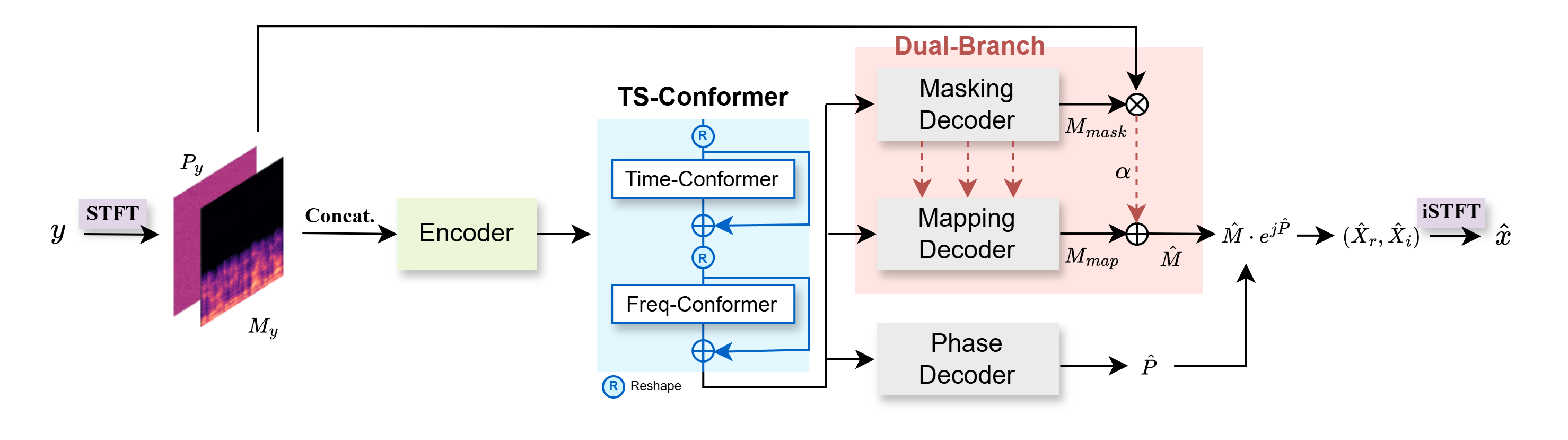}
  \caption{ Overall architecture of DBP-Net for general speech restoration. The model takes the STFT of a noisy input waveform and encodes the magnitude and wrapped phase into a compact representation. The dual-branch magnitude decoder performs noise/reverberation suppression via masking and high-frequency reconstruction via mapping, with parameter sharing across branches. A cross-branch skip fusion mechanism selectively transfers low-frequency content from the masking-based path to the mapping-based path. The final enhanced magnitude $\hat{M}_{\text{final}}$ is combined with the predicted phase $\hat{P}$ from the phase decoder to reconstruct the complex spectrum $(\hat{X}_{r}, \hat{X}_{i})$, which is used to compute both waveform- and spectrum-level losses via inverse STFT.}
  \label{fig:main}
\end{figure*}

\section{Model Description} \label{sec:proposed}

\subsection{Problem Formulation}

We perform general speech restoration in environments suffering from a combination of three common distortions: additive noise, reverberation, and bandwidth degradation. Let $x$ denote the clean speech signal. The observed distorted signal $y$ is simulated as:
\begin{equation}
y = h(x * r + n),
\end{equation}
where $r$ is a room impulse response convolved with $x$ to simulate reverberation, $n$ denotes additive noise, and $h(\cdot)$ is a low-pass filtering function that introduces bandwidth degradation by removing high-frequency components for digital transmission, respectively. The goal behind general speech restoration is to reconstruct a clean and perceptually enhanced speech signal $\hat{x}$ from the corrupted observation $y$.

This problem is quite challenging because denoising and dereverberation benefit from suppression or removal-based approaches, whereas bandwidth extension requires the generation of missing spectral content. To tackle these conflicting demands, we propose a unified architecture—termed the dual-branch parallel network (DBP-Net)—that explicitly decouples enhancement and generation processes within a joint framework.

\subsection{Overview of DBP-Net}

DBP-Net is designed to address general speech restoration by jointly performing noise and reverberation removal, as well as bandwidth reconstruction. The model processes the short-time Fourier transform (STFT) of the input signal and outputs enhanced magnitude and phase spectra, which are then used to reconstruct the time-domain waveform via inverse STFT.

The architecture consists of four main components:
\begin{itemize}
    \item A spectral encoder that transforms the magnitude and wrapped phase spectra into a compact time-frequency representation.
    \item A two-stage Conformer block that first models temporal and then spectral dependencies using convolution-augmented attention mechanisms.
    \item A dual-branch magnitude decoder that separately handles noise suppression and high-frequency generation.
    \item A phase decoder that estimates the clean phase spectrum.
\end{itemize}

A schematic diagram of the DBP-Net architecture is illustrated in Figure~\ref{fig:main}.

\subsection{DBP-Net Architecture}
\noindent\textbf{Spectral Input Representation}

We apply an STFT to the input waveform $y$ to obtain its complex-valued spectrogram. The magnitude and wrapped phase spectra, denoted as $M_{y}$ and $P_{y}$, are separately extracted and used as input features. Following prior work \cite{mpsenet}, the wrapped phase $P_{y}$ is represented using a sine-cosine embedding to handle discontinuities. These two components are concatenated along the channel dimension and passed to the encoder for joint processing of magnitude and phase information.

\noindent\textbf{Encoder and Transformer Backbone}

The encoder consists of multiple convolutional layers that downsample the input spectrogram while increasing the channel dimension. This results in a compressed time-frequency representation with a rich local structure.

To model both temporal and spectral dependencies, we adopt a two-stage Conformer-based backbone inspired by TS-Conformer~\cite{tsconformer}. Let $
D \in \mathbb{R}^{B \times T \times F \times C}$
denote the encoder output, where $B$, $T$, $F$, and $C$ represent the batch size, number of time frames, frequency bins, and channel dimension, respectively.
For temporal modeling, $D$ is reshaped into $D_T \in \mathbb{R}^{(B \cdot F) \times T \times C}$
by folding the frequency axis into the batch dimension. The reshaped representation is then processed by Conformer blocks that integrate multi-head self-attention and convolutional modules to capture long-range temporal dependencies. After temporal processing, the output is reshaped back to $\mathbb{R}^{B \times T \times F \times C}.$
For frequency modeling, the representation is further reshaped into
$ D_F \in \mathbb{R}^{(B \cdot T) \times F \times C}$
by folding the time axis into the batch dimension. A second set of Conformer blocks is applied to model spectral correlations across frequency bins. The final output is reshaped back to the original tensor size $B \times T \times F \times C$. This two-stage processing disentangles temporal and spectral dependencies, enabling the model to learn distortion-robust representations for both suppression and reconstruction tasks.

We use magnitude $M_y$ and wrapped phase $P_y$ as input features, concatenated along the channel dimension before being fed into the encoder. Phase wrapping effects are handled in the training objective via an anti-wrapping formulation following~\cite{mpsenet}.

\noindent\textbf{Dual-Branch Magnitude Decoder}

The magnitude decoder is split into two parallel branches: one for noise and reverberation removal, and the other for high-frequency reconstruction. Both decoders share the same architecture and parameters, except for their activation functions and interaction mechanisms.
\paragraph{\textit{Masking-based decoder} \textnormal{(denoising path)}} This branch adopts a masking-based strategy with a Sigmoid activation function to estimate a suppression mask for the distorted magnitude spectrum. This approach has been shown to be effective for denoising and dereverberation \cite{wang2018supervised}.

\paragraph{\textit{Mapping-based decoder} \textnormal{(BWE path)}} The second branch follows a mapping-based strategy with ReLU activation to directly predict a clean magnitude spectrum, particularly focusing on generating missing high-frequency content.

\noindent\textbf{Cross-Branch Skip Fusion}

Since direct skip connections from input to output are detrimental in multi-distortion settings—due to the risk of reintroducing noise—we propose a novel skip fusion strategy that selectively transfers enhanced low-frequency information from the denoising branch to the BWE branch.

Let $M_{\text{mask}}$ and $M_{\text{map}}$ denote the outputs of the masking-based and mapping-based decoders, respectively. We define the fused magnitude estimate $\hat{M}$ as:
\begin{equation}
\hat{M} = M_{\text{map}} + \alpha \cdot M_{\text{mask}},
\end{equation}
where $\alpha$ is a learnable parameter controlling the influence of the denoised low-frequency components on the generative path. This skip fusion enables the mapping-based decoder to incorporate enhanced spectral information without direct reliance on noisy input features.

\noindent\textbf{Final Output}

The final enhanced complex spectrogram is constructed by combining the fused magnitude estimate $\hat{M}$ with the predicted phase $\hat{P}$ as follows:
\begin{equation}
\hat{X} = \hat{M} \cdot e^{j\hat{P}}.
\end{equation}
Finally, the enhanced time-domain waveform $\hat{x}$ is obtained by applying the inverse STFT to $\hat{X}$.

\subsection{Training Objective}

We employ a multi-level loss function for training the proposed DBP-Net, targeting both waveform and spectral reconstruction.
Specifically, the training objective includes a time-domain $L_1$ loss:
\begin{equation}
\mathcal{L}_{\text{time}} = \| \hat{x} - x \|_1,
\end{equation}
In addition, magnitude and complex spectrum losses are computed in the time-frequency (TF) domain:

\begin{equation}
\mathcal{L}_{\text{mag}} = \| \hat{M} - M_{\text{gt}} \|_2^2, 
\end{equation}
\begin{equation}
    \mathcal{L}_{\text{com}} = \| \hat{X}_r - X_r \|_2^2 + \| \hat{X}_i - X_i \|_2^2,
\end{equation}

\noindent where $M_{\text{gt}}$ is the ground-truth magnitude spectrum, and $(X_r, X_i)$ denote the real and imaginary parts of the clean complex spectrum.

To further improve perceptual quality and phase accuracy, we additionally incorporate a metric-guided loss based on PESQ \cite{recommendation2001perceptual}, and a phase-aware loss based on anti-wrapping formulations, as proposed in \cite{mpsenet}. These auxiliary losses guide the model toward perceptually relevant and physically consistent spectral predictions.

The total loss is defined as a weighted combination:
\begin{equation}
\mathcal{L}_{\text{total}} = \gamma_1 \mathcal{L}_{\text{time}} + \gamma_2 \mathcal{L}_{\text{mag}} + \gamma_3 \mathcal{L}_{\text{com}} + \gamma_4 \mathcal{L}_{\text{metric}} + \gamma_5 \mathcal{L}_{\text{phase}},
\end{equation}
where all $\gamma$ weights are tunable hyperparameters that control the relative importance of each loss component.

\begin{table}[t]
\caption{Comprehensive experimental results for baseline models and step-wise integration process. For all metrics except LSD, higher values indicate better performance.}
\label{tab:main_results}
\centering
\resizebox{\linewidth}{!}{\begin{tabular}{cccccccccc}
\toprule
Method    &  CSIG ($\uparrow$) & CBAK ($\uparrow$) & COVL ($\uparrow$) &PESQ ($\uparrow$) & STOI ($\uparrow$)& SRMR ($\uparrow$) & LSD ($\downarrow$)  & \#Param. \\  \addlinespace[0.3mm] \hline \hline \addlinespace[0.3mm]
Noisy &     1.80 & 2.25 & 1.73 & 1.78 & 0.78 & 5.91 &  4.78 & -        \\ \addlinespace[0.3mm]\hdashline \addlinespace[0.3mm]
Baseline models & & & & & & & & & \\ \addlinespace[0.6mm]
VoiceFixer    & 3.21 & 2.12 & 2.54 & 1.82 & 0.83 & 9.13 &  2.72 & 122 M        \\
HD-DEMUCS   & 3.29 & 2.63 & 2.60 & 1.85 & 0.83 & 7.15 &  2.43 & 24 M       \\
SGMSE+  & 3.35 & 3.08 & 2.91 & 2.37 & 0.90 & \textbf{9.86}  & 2.88 & 65 M        \\ \addlinespace[0.3mm] \hdashline \addlinespace[0.3mm]
\textbf{Proposed} & & & & & & & & \\ \addlinespace[0.6mm]
DBP-Net   & \textbf{3.90}  & 3.11  & \textbf{3.31} & 2.61  & \textbf{0.92}  & 9.81  &  \textbf{2.24}  & \textbf{2.05 M}   \\     
  w/o skip fusion   & 3.79  & 3.05  & 3.25 & 2.60  & 0.91  & 9.59  & 2.37 & \textbf{2.05 M} \\
  w/o parameter sharing       & 2.74  & \textbf{3.14}  & 2.76 & \textbf{2.67} & 0.90  & 9.19  &  3.95 & 2.43 M       \\
\bottomrule
\end{tabular}}
\end{table}

\begin{figure*}[]
  \centering
  \includegraphics[width=0.95\linewidth]{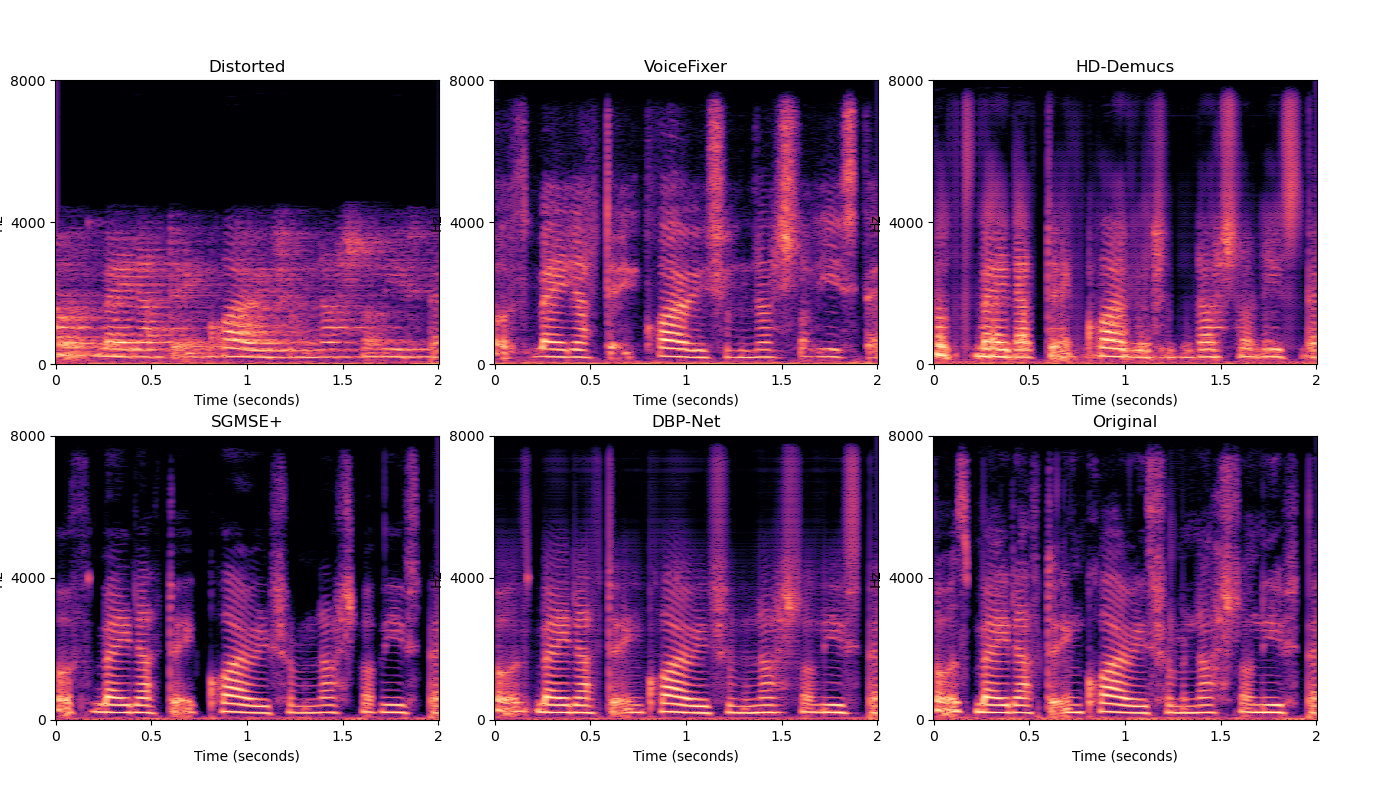}
  \caption{Comparison of spectrograms depicting the baselines and the proposed DBP-Net model. Generative models such as VoiceFixer and SGMSE+ produced more natural spectrograms, but the overall performance was inferior. The HD-DEMUCS model exhibited less effective noise and reverberation removal in the low-frequency band. Conversely, the proposed model demonstrated satisfactory noise and reverberation removal performance, along with effective bandwidth generation.}
  \label{fig:spectrogram}
\end{figure*}

\section{Experiments} \label{sec:experiments}
\subsection{Datasets and Experimental setup}
We considered noise, reverberation, and bandwidth degradation at a sampling rate of 16 kHz to generate distorted speech. For noise distortion, we mixed the VCTK corpus \cite{valentini2017noisy} consisting of 28 English speakers with the DEMAND noise dataset in a signal-to-noise ratio range of 0 – 20 dB.
The reverberant signal was generated by convolving the speech signal with a simulated room impulse response using a Pyroomacoustics engine. The dimensions of the reverberant room ranged from 5 to 10 m in length and width and 2 to 6 m in height, with the reverberation times (RT60) ranging from 0.3 to 0.9 seconds. For non-echoic speech, the target dry room had a fixed absorption coefficient of 0.99.
The bandwidth degradation signal was generated using various types of low-pass filters (e.g. Butterworth, Bessel, Chebyshev, and elliptic) with randomly selected cut-off frequencies between 2 and 4 kHz.
We selected ``p258" and ``p287" speakers from the training set for the validation set. 

We trained the proposed DBP-Net in Figure~\ref{fig:main} for 1M steps using the AdamW optimizer~\cite{loshchilov2017decoupled} with a learning rate of 0.0005. 
The contributions of the dual-branch magnitude decoders were balanced by a learnable skip-fusion parameter $\alpha$, which was initialized at training onset and converged to approximately 0.38.
For the loss function, we adopted the weighting coefficients $\gamma_1$ to $\gamma_5$ from the MP-SENet training setup~\cite{mpsenet}, where the values were set to 0.2, 0.9, 0.1, 0.05, and 0.3, respectively.

\subsection{Evaluation Metrics}
We evaluated the speech restoration performance using various assessment metrics. 
For comprehensive assessments of speech performance, we analyzed the speech signal distortion, background noise, and quality using CSIG, which is the mean opinion score predictor of signal distortion, CBAK, which is the mean opinion score predictor of background-noise intrusiveness, and COVL, which is the mean opinion score predictor of overall signal quality \cite{composite}.
In addition, the speech quality and intelligibility were measured using wide-band perceptual evaluation of speech quality (PESQ) \cite{recommendation2001perceptual} and short-time objective intelligibility (STOI) \cite{taal2011algorithm}. To assess the effectiveness of speech dereverberation, we employed the speech-to-reverberation modulation energy ratio (SRMR) metric \cite{falk2010temporal}. For BWE, we used the log-spectral distance (LSD) to measure the performance. For all metrics except for the LSD, higher values indicate better performance.

\subsection{Comparison with Baseline Models}
Table~\ref{tab:main_results} presents a performance comparison between the proposed DBP-Net and existing baseline models, including VoiceFixer~\cite{liu2022voicefixer}, HD-DEMUCS~\cite{kim2023hd}, and SGMSE+~\cite{richter2023speech}. VoiceFixer and HD-DEMUCS are representative methods for general speech restoration, while SGMSE+ was originally proposed for speech denoising but also supports dereverberation and bandwidth extension~\cite{sgmse+bwe}. We used the official pretrained model for VoiceFixer\footnote{\url{https://github.com/haoheliu/voicefixer}}, reimplemented HD-DEMUCS using the official open-source code\footnote{\url{https://github.com/facebookresearch/denoiser.git}}, and trained SGMSE+ using its official repository\footnote{\url{https://github.com/sp-uhh/sgmse.git}}.

The proposed DBP-Net significantly outperformed all baselines across various metrics. In terms of perceptual quality, DBP-Net achieved the highest CSIG score of 3.90 and COVL of 3.31, indicating superior restoration of speech quality and overall naturalness.
This trend is also visually supported in Figure~\ref{fig:spectrogram}, where DBP-Net demonstrates effective noise suppression in low-frequency regions and accurate restoration of high-frequency components in a balanced manner.
In contrast, the best-performing baseline, SGMSE+, achieved CSIG and COVL scores of 3.35 and 2.91, respectively. For speech intelligibility, DBP-Net also recorded the highest STOI score of 0.92, outperforming SGMSE+ (0.90), HD-DEMUCS (0.83), and VoiceFixer (0.83).

Moreover, our model demonstrated the most effective bandwidth restoration performance, achieving the lowest LSD of 2.24, while the LSDs for SGMSE+, HD-DEMUCS, and VoiceFixer were 2.88, 2.43, and 2.72, respectively. Notably, DBP-Net achieved this superior performance with only 2.05 million parameters, which is substantially fewer than VoiceFixer (122M), HD-DEMUCS (24M), and SGMSE+ (65M). This highlights the efficiency of our architecture in both performance and model size.

\subsection{Ablation Study}
We conducted an ablation study to validate the effectiveness of each component in the proposed architecture. Starting from a basic dual-branch magnitude decoder setup, we progressively integrated parameter sharing and skip fusion to evaluate their individual contributions to overall performance.
We first examined a naive dual-decoder model without any parameter sharing between the two magnitude decoders. While this model achieved the highest PESQ score (2.67) and competitive CBAK (3.14), it failed to perform well in terms of CSIG (2.74), COVL (2.76), and LSD (3.95), indicating that bandwidth restoration was insufficient and speech quality remained degraded. This result suggests that training the two decoders independently hindered the model's ability to generalize across multiple distortions.

The elevated PESQ score in this variant reflects strong suppression performance but does not imply accurate full-band reconstruction. While PESQ is largely influenced by perceptual similarity in dominant frequency regions, LSD directly evaluates spectral distortion across the entire frequency band. The substantially higher LSD value (3.95) therefore indicates insufficient high-frequency reconstruction despite competitive suppression metrics. This observation underscores a key challenge in multi-distortion restoration: optimizing suppression-oriented metrics alone cannot ensure full-spectrum fidelity.

To address this limitation, we applied parameter sharing between the two decoders. This significantly improved LSD (from 3.95 to 2.37), as well as CSIG and COVL, demonstrating that joint learning allowed better utilization of both masking-based and mapping-based paths. The shared structure enabled effective bandwidth generation without sacrificing denoising capability.
Lastly, we introduced a learnable skip fusion mechanism to allow interaction between the two branches. This yielded further improvements in CSIG (from 3.79 to 3.90), LSD (from 2.37 to 2.24), and overall speech quality (COVL 3.31). Notably, this was achieved with a compact model size of only 2.05M parameters.
The results confirm that both parameter sharing and skip fusion contribute significantly to general speech restoration. The final proposed model, DBP-Net, outperformed all ablation variants and prior baselines across most metrics, achieving strong performance in noise and reverberation removal, as well as high-frequency bandwidth generation.

\section{Conclusion}
\label{sec:majhead}
In this letter, we proposed DBP-Net, a general speech restoration model that integrates two parallel magnitude decoders: one dedicated to distortion removal and the other to speech reconstruction. Motivated by the limitations of existing single-purpose models, we designed a unified architecture capable of addressing multiple distortions—such as noise, reverberation, and bandwidth reduction—within a single framework. The proposed model demonstrated consistent improvements across various objective metrics, achieving effective and compact speech restoration in challenging acoustic environments.


\bibliographystyle{cas-model2-names}

\bibliography{cas-refs}



\end{document}